\documentclass[aps,prc,twocolumn,amsmath,amssymb,superscriptaddress,nofootinbib,showpacs,showkeys]{revtex4}
\usepackage{dcolumn}
\usepackage{graphicx,color}
\usepackage{multirow}
\usepackage{textcomp}
\usepackage{physics}
\usepackage{dsfont}
\usepackage{epstopdf}
\usepackage{relsize}
\usepackage{tikz}	
\begin{document}
  \newcommand {\nc} {\newcommand}
  \nc {\beq} {\begin{eqnarray}}
  \nc {\eeq} {\nonumber \end{eqnarray}}
  \nc {\eeqn}[1] {\label {#1} \end{eqnarray}}
  \nc {\eol} {\nonumber \\}
  \nc {\eoln}[1] {\label {#1} \\}
  \nc {\ve} [1] {\mbox{\boldmath $#1$}}
  \nc {\ves} [1] {\mbox{\boldmath ${\scriptstyle #1}$}}
  \nc {\mrm} [1] {\mathrm{#1}}
  \nc {\half} {\mbox{$\frac{1}{2}$}}
  \nc {\thal} {\mbox{$\frac{3}{2}$}}
  \nc {\fial} {\mbox{$\frac{5}{2}$}}
  \nc {\la} {\mbox{$\langle$}}
  \nc {\ra} {\mbox{$\rangle$}}
  \nc {\etal} {\emph{et al.}}
  \nc {\eq} [1] {(\ref{#1})}
  \nc {\Eq} [1] {Eq.~(\ref{#1})}
  \nc {\Refc} [2] {Refs.~\cite[#1]{#2}}
  \nc {\Sec} [1] {Sec.~\ref{#1}}
  \nc {\chap} [1] {Chapter~\ref{#1}}
  \nc {\anx} [1] {Appendix~\ref{#1}}
  \nc {\tbl} [1] {Table~\ref{#1}}
  \nc {\Fig} [1] {Fig.~\ref{#1}}
  \nc {\ex} [1] {$^{#1}$}
  \nc {\Sch} {Schr\"odinger }
  \nc {\flim} [2] {\mathop{\longrightarrow}\limits_{{#1}\rightarrow{#2}}}
  \nc {\textdegr}{$^{\circ}$}
  \nc {\inred} [1]{\textcolor{red}{#1}}
  \nc {\inblue} [1]{\textcolor{blue}{#1}}
  \nc {\IR} [1]{\textcolor{red}{#1}}
  \nc {\IB} [1]{\textcolor{blue}{#1}}
  \nc{\pderiv}[2]{\cfrac{\partial #1}{\partial #2}}
  \nc{\deriv}[2]{\cfrac{d#1}{d#2}}
\title{Simplified dynamical eikonal approximation}
\author{C.~Hebborn}
\email{chloe.hebborn@ulb.ac.be}
\affiliation{Physique Nucl\'eaire et Physique Quantique (CP 229), Universit\'e libre de Bruxelles (ULB), B-1050 Brussels}
\author{D.~Baye}
\email{daniel.baye@ulb.ac.be}
\affiliation{Physique Nucl\'eaire et Physique Quantique (CP 229), Universit\'e libre de Bruxelles (ULB), B-1050 Brussels}
\date{\today}
\begin{abstract}
\begin{description}
\item[Background] Breakup reactions are often used to probe the nuclear structure of halo nuclei. The eikonal model diverges for Coulomb breakup since it relies on the adiabatic approximation. To correct this weakness, a Coulomb-corrected eikonal method (CCE) using the Coulomb first-order-perturbation approximation was developed.
\item[Purpose] Since the CCE mixes two reaction models and treats the Coulomb and nuclear interactions on different footings, we study here  an alternative approach. We develop a simplification to the dynamical eikonal approximation (S-DEA) which has a similar numerical cost as the usual eikonal model, and study its efficiency for both nuclear- and Coulomb-dominated breakup reactions. 
\item[Methods] {We compare the energy  and parallel-momentum cross sections obtained with the dynamical	eikonal approximation, the usual eikonal approximation, the CCE and the S-DEA.}
\item[Results] The S-DEA leads to  precise energy distributions for both breakup reactions. The corresponding parallel-momentum distributions obtained with the S-DEA are improved compared to the ones computed with the eikonal model. It is more efficient for nuclear-dominated breakup than the CCE since it reproduces better the shape and magnitude of the distribution.  However, for the Coulomb breakup, the distribution lacks asymmetry. 
\item[Conclusions] The simplification of the DEA developed in this work improves  significantly  the eikonal descriptions of  breakup energy distribution for both Coulomb- and nuclear-dominated reactions. The asymmetry of the parallel-momentum distribution  is enhanced for nuclear-dominated breakup. This study  confirms that the asymmetry is due to dynamical effects. A direct prospect of this work would be to extend this model to two-neutron halo-nucleus projectiles.
\end{description}

\end{abstract}
\maketitle
%


\section{Introduction}\label{Introduction}

 Halo nuclei are among the most peculiar nuclear structures~\cite{T96,HJ87}. Indeed, they exhibit a unusually large size compared to stable nuclei. This  size is due to the low binding of one or two nucleons, which tunnel far from the rest of the nucleons and form a diffuse halo around them. Accordingly, one can model these nuclei as few-body objects: a compact core with one or two neutrons in the halo.
 An archetypical halo nucleus is $^{11}\rm Be$, seen as a $^{10}\rm Be$ core with one loosely-bound neutron  in its halo.

 Due to their  short lifetime, halo nuclei cannot be studied through usual spectroscopic techniques, i.e., where they are fixed targets, but are probed through indirect techniques, such as reactions. Breakup reactions describe the dissociation of the halo from the core. Because the core-halo binding is  fragile, breakup observables can have  high   statistics and are often used in the  low-intensity exotic beam facilities~\cite{P03,Fetal04}. These reactions reveal the cluster structure inside the nucleus and, when dominated by Coulomb, can be used to extract cross sections of astrophysical interest~\cite{BBR86}. To infer reliable information about the nuclear structure, one needs  an accurate reaction model coupled to a realistic description of the projectile.

Very precise models have been developed to this aim (see a recent review in Ref.~\cite{BC12}): the continuum-discretized coupled-channels method (CDCC)~\cite{Kam86,YOMM12}, time-dependent models~\cite{BW81,AW75} and the  dynamical eikonal approximation (DEA)~\cite{BCG05}. These models are  time consuming and are thus often restricted to a simple description of the projectile, i.e., usually a two-body description of the projectile. Extensions to three-body projectiles have been developed, but they are often restricted to elastic scattering observables~\cite{RGetal09,FG15}. Simpler methods, such as the usual eikonal model~\cite{G59}, are cheaper from a computational viewpoint and provide a simple interpretation of the collision. The  eikonal model relies on an adiabatic approximation and  is thus valid at energies above 60$A$~MeV, where breakup reactions are usually measured. However, its adiabatic treatment of the collision has three main drawbacks:  (i) its wave functions do not have the right asymptotic behavior, (ii) it  exhibits an additional symmetry that more elaborate methods do not have, and (iii) it is incompatible with the Coulomb interaction, causing a divergence in the breakup observables.

 Since the eikonal model provides precise results for nuclear-dominated reactions, a correction to its inefficient treatment of the Coulomb interaction -- called the Coulomb-corrected eikonal model (CCE)~\cite{MBB03,CBS08} -- has been developed. This correction replaces the diverging term of the eikonal phase by  the first-order approximation of the perturbation theory~\cite{AW75}.    In Ref.~\cite{CBS08}, the authors show that the CCE provides accurate energy and parallel-momentum distributions for the Coulomb-dominated breakup of one-neutron halo nucleus. Its extension to  two-neutrons halo nuclei  also leads to precise Coulomb breakup cross section~\cite{BCDS09,PDB12}. Nonetheless, this correction mixes two reaction models and   treats the nuclear and the Coulomb interactions on different footings.

 In this work, we develop a simplification to the DEA (S-DEA) which naturally removes the Coulomb divergence and {partly }corrects the asymptotic behavior, i.e.\ it tends to the first-order of  the perturbation theory.  This model is derived from an approximate solution of the DEA equation and has the advantage of having a numerical cost similar to the usual eikonal model. In the present article, we study the efficiency of this model  in the case of the nuclear- and Coulomb-dominated breakups of a one-neutron halo nucleus; we consider $^{11}\rm Be$ impinging on  $^{12}\rm C$ and $^{208}\rm Pb$ targets at 67$A$~MeV and 69$A$~MeV, respectively. These reactions were measured at RIKEN~\cite{Fetal04} and are precisely reproduced by the full DEA calculations~\cite{GBC06}. We thus compare our S-DEA  results to the DEA and to the CCE.

 Sec.~\ref{Sec2} describes the reaction model and provides the main ideas of the DEA, the CCE and the S-DEA. In Sec.~\ref{Sec3}, we evaluate the accuracy of the S-DEA for the breakup distribution of $^{11}\rm Be$ as a function of the $^{10}\rm Be$-$n$ relative energy and parallel momentum. This analysis is made for both  nuclear- and Coulomb-dominated breakups. Sec.~\ref{Conclusions} summarizes the main conclusions of this work.

\section{Theoretical framework}\label{Sec2}
	\subsection{Reaction model}\label{Sec2A}
 We study the  breakup of a one-neutron halo-nucleus projectile $P$ of charge $Z_P e$ with a structureless target $T$ of mass $m_T$ and charge $Z_T e$. Since halo nuclei exhibit a  clusterized structure, we model $P$ as a two-body object, composed of a core $c$ of mass $m_c$ and charge $Z_Pe$, and a neutron $n$ of mass $m_n$. We assume all three particles structureless and only the spin of the neutron is considered. The structure of the projectile is described through an effective internal Hamiltonian~\cite{BC12}
	\begin{equation}
h_{cn} = \frac{p^2 }{2 \mu_{cn}}  + V_{cn}(r),\label{eq1}
\end{equation}
where $\ve{p}$ and $\ve{r}=(\ve{s},z)$ are respectively the $c$-$n$ relative momentum and coordinate, $\mu_{cn}$ is the $c$-$n$ reduced mass and $V_{cn}$ is an effective $c$-$n$ potential, that we assume local. Here, this real potential is taken as the sum of a Woods-Saxon and a spin-orbit terms (see Sec.~\ref{Sec3}).
 
 The interactions of both fragments $c$ and $n$   with the target $T$ are simulated through central local optical potentials   $V_{cT}$ {(including Coulomb)} and $V_{nT}$, respectively. The collision is thus described by the three-body \Sch equation~\cite{BC12}
	\begin{eqnarray}
	\lefteqn{\left[\frac{P^2}{2\mu}+h_{cn}+V_{cT}(R_{cT})+V_{nT}(R_{nT})\right]\Psi(\ve{R},\ve{r})=}\nonumber\\
	&\hspace{5cm}&E_{\rm tot}\ \Psi(\ve{R},\ve{r}), \label{eq2}
	\end{eqnarray}
	where $\ve{P}$ and $\ve{R}=(\ve{b},Z)$ are respectively the $P$-$T$ relative momentum and coordinate, $\mu$ is the $P$-$T$ reduced mass, $E_{\rm tot}$ is the total energy of the system and $\ve{R_{(c,n)T}}=(\ve{b_{(c,n)T}},Z_{(c,n)T})$ are respectively the $c$-$T$ and $n$-$T$ relative coordinates. 
This \Sch equation is solved with the initial condition that the projectile is in its ground state $\phi_0$ with energy $E_0$, and is impinging on the target with a velocity $\ve{v}=\ve{P}/\mu=\hbar \ve{K}/\mu$. The wave function verifies 
		\begin{equation}
	\Psi(\ve{R},\ve{r})\flim{Z}{- \infty}e^{iKZ+\cdots}\ \phi_{0} (\ve{r}),\label{eq3}
	\end{equation}
	where we choose the $Z$-axis to be the beam axis.	The ``$\cdots$" in Eq.~\eqref{eq3} reflects the fact that the long-range Coulomb interaction distorts the incoming plane wave, even at large distances.

	\subsection{Dynamical eikonal approximation}\label{Sec2B}

	 At high enough energy, the beam is only slightly deflected and the three-body wave function $\Psi$ does not differ much from a plane wave. The eikonal model~\cite{G59} factorizes this plane wave out of the wave function
		\begin{equation}
	\Psi(\ve{R},\ve{r}) =e^{iKZ}\ \widehat{\Psi}(\ve{R},\ve{r})\label{eq4}
	\end{equation}
	and assumes that the second derivatives of the new wave function $\widehat{\Psi}$  can be neglected. The three-body \Sch equation~\eqref{eq2} simplifies into the DEA equation~\cite{BCG05}  
			\begin{eqnarray}
	\lefteqn{	i\hbar v \pderiv{}{Z} \widehat{\Psi}^{\rm DEA}(\ve{R},\ve{r})=}\nonumber\\
		&\hspace{0cm}&[h_{cn}-E_0+V_{cT}(R_{cT})+V_{nT}(R_{nT})]\widehat{\Psi}^{\rm DEA}(\ve{R},\ve{r}).\label{eq5}
	\end{eqnarray}	
 This partial-derivative equation is solved through a numerical evolution calculation as a function of the transverse coordinate \ve{b} \cite{BCG05}. Its solutions {at large $b$s} tend  asymptotically to the {ones obtained at the perturbation theory}  and are  accurate at energies above 40$A$~MeV~\cite{GBC06,GCB07}.

		\subsection{Eikonal approximation}\label{Sec2C}

 The usual eikonal model~\cite{G59} relies on the adiabatic approximation, which sees the coordinate of the projectile as frozen during the collision. Consequently, the internal Hamiltonian {is approximated} by the energy of the ground state  in Eq.~\eqref{eq5}, i.e. $h_{cn}\approx E_0$. The solutions satisfying \Eq{eq4} can be derived analytically and have the simple asymptotic form~\cite{BC12}
	\begin{eqnarray}
\widehat{\Psi}^{\rm eik}(\ve{R},\ve{r})&\flim{Z}{+\infty}&e^{i \chi_{cT}(\ve{b},\ve{s})+ i\chi_{nT}(\ve{b},\ve{s})} \phi_{0}(\ve{r})\nonumber\\
&=& e^{i\chi^C_{PT}(b)+i\chi^C(\ve{b},\ve{s}) +i \chi^N(\ve{b},\ve{s})}\phi_{0}(\ve{r}),\label{eq6}
\end{eqnarray}
with the eikonal phases  defined as
\begin{equation}
\chi_{(c,n)T}(\ve{b},\ve{s})=-\frac{1}{\hbar v} \int^{+\infty}_{-\infty} V_{(c,n)T} (\ve{b_{(c,n)T}},Z)\ \mathrm{d}Z.\label{eq7}
\end{equation}
The phases $\chi^N$ corresponds to the nuclear  potential and the Coulomb phases  $\chi^C$ and $\chi^C_{PT}=2\eta \ln (Kb)$ to the Coulomb tidal force and the $P$-$T$ Coulomb scattering, respectively.
 The eikonal solutions~\eqref{eq6} can be interpreted semiclassically as $P$ following a straight-line trajectory and accumulating a phase through its interaction with $T$ during the collision. The main advantage of this model is that  the whole information about the collision is contained within the eikonal phases. 
 This model is restricted to high energies, i.e.\ above 60$A$~MeV, where straight-line trajectories make sense.

  Compared to more elaborate models, e.g.\ CDCC and the DEA, the eikonal model exhibits an additional symmetry across the plane of the projectile's internal transverse  coordinate~$\ve{s}$~\cite{BCDS09}. Consequently, the breakup distributions as a function of the relative $c$-$n$ parallel-momentum    are symmetric. One can also note that the eikonal solutions~\eqref{eq6} {at large $b$s} do not tend asymptotically to the {ones obtained with the perturbation theory}. Moreover,  the adiabatic approach is incompatible with the infinite range of the Coulomb interaction. The eikonal phases are therefore not well defined for this interaction and the eikonal breakup matrix elements diverge~\cite{CBS08}. 
 
		\subsection{Coulomb-corrected eikonal approximation}\label{Sec2D}

   The CCE~\cite{MBB03} was developed to remove the divergence from the breakup matrix element. The idea is to replace within the matrix element the diverging  Coulomb first-order term $\chi^C$ by the Coulomb first-order-perturbation approximation $\chi^C_{{\rm FO}}$~[see Eq.~{~\eqref{eq13}} below]~\cite{MBB03,CBS08}
 \begin{eqnarray}
e^{i\chi^C_{PT}} e^{i\chi^C}  e^{i\chi^N} \to e^{i\chi^C_{PT}}\left(e^{i\chi^C}-i\chi^C +i\chi^C_{{\rm FO}}\right) e^{i\chi^N}.	\label{eq8}
 \end{eqnarray}
 In the CCE~\cite{MBB03,CBS08}, $\chi_C^{{\rm FO}}$ is approximated by its dipole contribution, which is dominant for $^{11}\rm Be$~\cite{CB05}. 
Beside removing the divergence, this model  {partly} corrects the asymptotic  behavior {at large $b$s} and does not exhibit the additional symmetry of the eikonal model. Nevertheless,  the nuclear and Coulomb interactions are not treated on the same footing.

	\subsection{Simplified dynamical eikonal approximation}\label{Sec2E}

In this work, we develop a simplification to the DEA, combining the simplicity of the usual eikonal model while removing its divergence and  treating both interactions within a unique framework. This model is based on a unitary transformation of the wave function $\widehat{\Psi}$ of Eq.~\eqref{eq4}
			\begin{eqnarray}
	\widetilde{\Psi}(\ve{R},\ve{r}) &=&e^{\frac{i}{\hbar v}\left[\left(h_{cn}-E_0\right){Z}+\int^Z_{-\infty} V^C_{PT}(R') dZ'\right]} \widehat{\Psi}(\ve{R},\ve{r}),\label{eq9}
	\end{eqnarray}
	where $V^C_{PT}$ is the $P$-$T$ Coulomb interaction. 
The three-body \Sch equation~\eqref{eq5} thus becomes
	\begin{eqnarray}
	  \lefteqn{	i\hbar v \pderiv{}{Z} \widetilde{\Psi}(\ve{R},\ve{r})=}\nonumber \\
	&&\hspace{-0.3cm}e^{\frac{i}{\hbar v}h_{cn}Z}[V_{cT}(R_{cT})+V_{nT}(R_{nT})-V^C_{PT}(R)]e^{-\frac{i}{\hbar v}h_{cn}Z}\nonumber \\
	&&\hspace{1cm}\times\widetilde{\Psi}(\ve{R},\ve{r}). \label{eq10}
	\end{eqnarray}
Note that Eq.~\eqref{eq10} is equivalent to the DEA equation~\eqref{eq5} and is at the basis of all perturbation treatments~\cite{Joa75}. 
	
	Here, we adopt another strategy, we  approximate the solution {of Eq.~\eqref{eq10} by only keeping the first term in the exponent of its Magnus expansion or, equivalently, the first factor of its Fer expansion} \cite{M54,F58,W67}
	\begin{eqnarray}
 		\lefteqn{\widetilde{\Psi}(\ve{R},\ve{r})\approx}\nonumber \\
 		&& e^{-\frac{i}{\hbar v}\int_{-\infty}^{Z} e^{\frac{i}{\hbar v}h_{cn}Z'}[V_{cT}(R'_{cT})+V_{nT}(R'_{nT})-V^C_{PT}(R')]e^{-\frac{i}{\hbar v}h_{cn}Z'}dZ'}\nonumber \\
 		&& \hspace{1cm}\times\phi_{0}(\ve{r}),\label{eq11}
\end{eqnarray}	
where $R'_{cT}$, $R'_{nT}$ and $R'$ depend on $Z'$. {Asymptotically, this new wave function tends to the ground state of the projectile, i.e., $\widetilde{\Psi}(\ve{R},\ve{r}) \flim{Z}{- \infty} \phi_0(\ve{r})$.} With this approximation, we neglect all the terms containing commutators, appearing in higher-orders terms of both Magnus and Fer expansions.

To compute the breakup matrix element, we make a second approximation; we replace the operators $h_{cn}$ in the exponentials~\eqref{eq11} by the eigenvalues corresponding to the closest wave function, i.e., the final energy $E$ on the left-hand side and the initial energy $E_0$ on the right-hand side. We therefore obtain an eikonal-like model where the exponential of the eikonal phases within the breakup matrix element is replaced according to 
\begin{eqnarray}
e^{i\chi^C_{PT}} e^{i\chi^C} e^{i\chi^N} \to e^{i\chi^C_{PT}}e^{i\chi^C_{{\rm FO}}}e^{i\chi^N_{{\rm FO}}}\label{eq12}
\end{eqnarray}
with $\chi^{(C,N)}_{{\rm FO}}$ the first-order-perturbation approximations of respectively the Coulomb and nuclear interactions~\cite{AW75}. 

 The Coulomb part is evaluated analytically~\cite{AS64}
 \begin{eqnarray}
\chi^C_{{\rm FO}}(\ve{b},\ve{r})=& -\eta \int_{-\infty}^{+\infty}e^{i\frac{\omega Z}{v}}\left(\frac{1}{R_{cT}}-\frac{1}{R}\right) dZ\label{eq13}\\
=&-2\eta \left[e^{i\frac{\omega}{v}\frac{m_n}{m_P}z}K_0\left(\frac{\omega}{v}b_{cT}\right)-K_0\left(\frac{\omega}{v}b\right)\right]\label{eq14}
\end{eqnarray}
where $\omega=\frac{E-E_0}{\hbar}$, $m_P=m_n+m_c$, and $\eta= Z_TZ_P e^2/(4\pi \epsilon_0 \hbar v)$ is the $P$-$T$ Sommerfeld parameter. Compared to the CCE which only uses the dipole contribution, we consider the whole Coulomb first-order-perturbation  approximation. However, as previously mentioned, for $^{11}\rm Be$, the higher multipoles do not contribute significantly.

{We rewrite the nuclear part of the first-order-perturbation approximation as}
\begin{eqnarray}
\chi^N_{{\rm FO}}(\ve{b}, \ve{r})&=& - \frac{1}{\hbar v}  \int_{-\infty}^{+\infty}e^{ i\frac{\omega Z}{v}}
\Bigg\{
e^{-i \frac{\omega}{ v } \frac{m_c}{m_P} z}  V_{nT}\left(\sqrt{b_{nT}^2+Z^2}\right)\nonumber\\
&&\hspace{-1.6cm}+e^{i \frac{\omega}{ v} \frac{m_n}{m_P}z} \bigg[V_{cT}\left(\sqrt{b_{cT}^2+Z^2}\right)-\frac{\hbar v\eta}{\sqrt{b_{cT}^2+Z^2}}\bigg] \Bigg\}dZ\label{eq15}
\end{eqnarray}
and we compute this integral  numerically. 
 Additionally to its simple implementation, the S-DEA wave functions  naturally tend asymptotically to the first-order of the perturbation theory {at large $b$s}. Moreover, this model does not introduce the additional symmetry of the eikonal model across the plane defined by $\ve{s}$.

{Approximation \Eq{eq12} has a serious drawback, caused by the imaginary part of the first-order-perturbation approximations~\eqref{eq14}--\eqref{eq15}. When the imaginary part of the first-order-perturbation approximation is negative, the cross sections may take unrealistic values 	as $\exp[i\chi^C_{\rm FO}+i\chi^N_{\rm FO}]$ then involve increasing real exponentials. 
This problem is   enhanced by  the use of optical potentials. 	Such potentials contain an imaginary part simulating the absorption into channels other than the elastic one in the core or neutron interaction with the target.   Consequently, the Hamiltonian is not Hermitian and the $S$-matrix looses its unitarity. Even for purely real potentials, an imaginary part would arise from the imaginary exponential multiplying the potentials. }
{To partly cure   this problem, we treat separately the absorptive part of the potentials with the usual eikonal approximation. This ensures that the imaginary parts of the optical potentials suppress the unphysical contributions.  The S-DEA model is thus defined by }
\begin{eqnarray}
 e^{i\chi^C_{PT}}  e^{i\chi^C} e^{i\chi^N} \to e^{i\chi^C_{PT}}  e^{i\chi^C_{{\rm FO}}}e^{i\chi^N_{{\rm S-DEA}}}\label{eq16}
\end{eqnarray}
{with}
\begin{eqnarray} 
\chi^N_{{\rm S-DEA}}(\ve{b}, \ve{r}) = \mathrm{Im}\, \chi^N(\ve{b}, \ve{r}) \nonumber \\
- \frac{1}{\hbar v}  \int_{-\infty}^{+\infty}e^{ i\frac{\omega Z}{v}}
\Bigg\{e^{-i \frac{\omega}{ v } \frac{m_c}{m_P} z}{ \rm Re} \,V_{nT}\left(\sqrt{b_{nT}^2+Z^2}\right) \nonumber \\  
+e^{i \frac{\omega}{ v} \frac{m_n}{m_P}z} \bigg[{ \rm Re}\,V_{cT}\left(\sqrt{b_{cT}^2+Z^2}\right)-\frac{\hbar v\eta}{\sqrt{b_{cT}^2+Z^2}}\bigg]
\Bigg\}dZ\label{eq17}
\end{eqnarray}
where $\chi^N$ has been defined in {Sec.}~\ref{Sec2C}. 

\section{Results}\label{Sec3}

	 In this article, we study the S-DEA with the breakup of $^{11}\rm Be$ on $^{12}\rm C$  and  $^{208}\rm Pb$ targets. We use the same description of $^{11}\rm Be$ as in Refs.~\cite{CGB04,CBS08}:  it is seen  as  $^{10}\rm Be$ core in its $0^+$ ground state to which a neutron is bound by $0.504$~MeV. The $^{10}\rm Be$-$n$ interaction  is simulated by a Woods-Saxon potential and a spin-orbit term,  adjusted to the three first levels: $1/2^+$, $1/2^-$ and $5/2^+$, modelled respectively as a $1s1/2$ state,  a $0p1/2$ state and  a $d5/2$ resonance. The parameters of this $^{10}\rm Be$-$n$ potential are given in Ref.~\cite{CGB04}. The $^{10}\rm Be$-$T$ and $n$-$T$ interactions are simulated through the same optical potentials as in Ref.~\cite{CBS08}. Our eikonal, CCE and DEA calculations have the same numerical inputs  as in   Ref.~\cite{CBS08}. The numerical computations of the S-DEA  use the same meshes as  the CCE.
	 
	 \begin{figure*}
	 	\center
	 		 	{\includegraphics[width=0.45\linewidth]{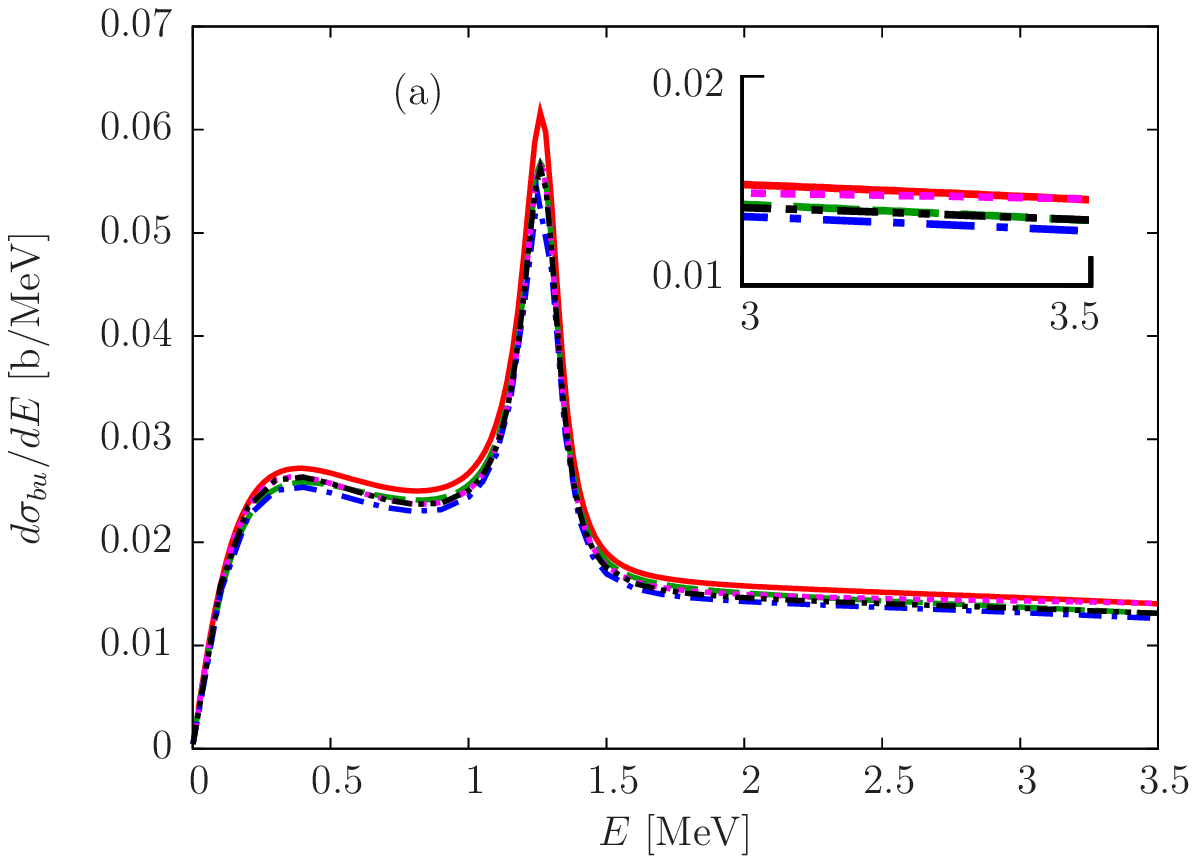}}
	 	\hspace{0.5cm}
	 	{\includegraphics[width=0.45\linewidth]{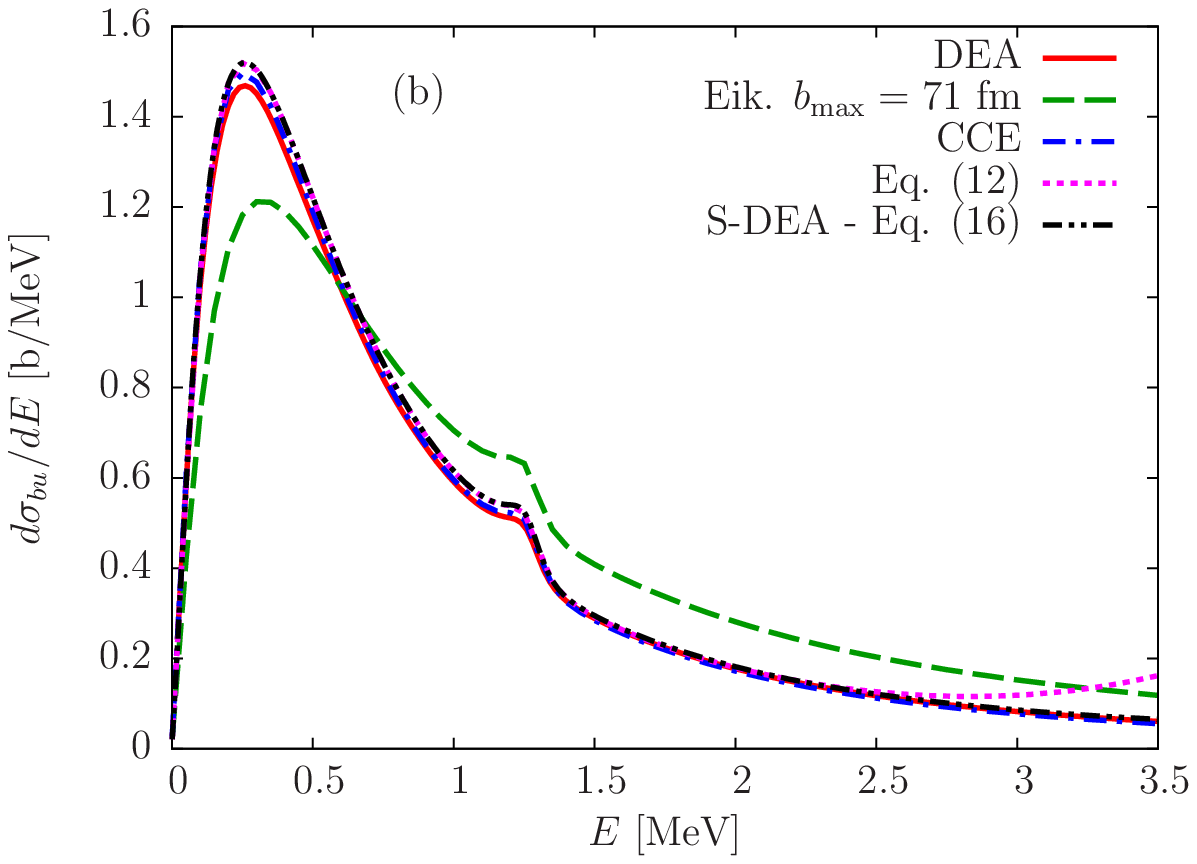}}	
	 	\caption{Diffractive breakup cross sections of $^{11}\rm Be$ with  (a) $^{12}\rm C$ at $67A$~MeV and (b) $^{208}\rm Pb$ at 69$A$~MeV  as a function of the  $^{10}\mathrm{Be}$-$n$ relative energy $E$. In  panel (a), we also plot a zoom of the energy distributions between 3~MeV and 3.5~MeV. } \label{FigSbudE}
	 \end{figure*}
To evaluate the accuracy of the S-DEA, we compute the breakup cross section as a function of the relative $c$-$n$ energy, displayed in Fig.~\ref{FigSbudE}. The panel (a) corresponds to a  nuclear-dominated breakup,  $^{11}\rm Be$ on $^{12}\rm C$ at $67A$~MeV, and the panel (b) to a Coulomb-dominated reaction, $^{11}\rm Be$ on $^{208}\rm Pb$ at $69A$~MeV. For these two reactions, the DEA (solid red lines) reproduces well the RIKEN data~\cite{Fetal04,BCG05,GBC06} and is used as reference. As previously mentioned, the eikonal model (dashed green lines) does not treat properly the Coulomb interaction and therefore requires the use of an upper cutoff $b_{\rm max}$ to provide breakup calculations~\cite{AS00}. With this cutoff, the usual eikonal model leads to results close to the DEA ones for the nuclear-dominated reaction but fails to describe the Coulomb-dominated breakup, due to the incompatibility of this long-range interaction with the adiabatic assumption.

Both panels (a) and (b) of Fig.~\ref{FigSbudE} show that the CCE (dash-dotted blue lines) improves the eikonal treatment of the Coulomb interaction and gives accurate energy distributions for both reactions. Let us first consider the approximation defined by \Eq{eq12}. For the nuclear-dominated breakup distribution (a), it leads to exactly the same results (dotted magenta lines) as the CCE at $E<1.5$~MeV. The Coulomb-dominated cross section (b) shows that it is accurate at $E<2.5$~MeV. Indeed, even if it slightly overestimates the peak of the distribution, it lies close to the DEA and the CCE. For both reactions, the approximation \eqref{eq12} starts to increase at higher energies and completely fails to reproduce the shape and the magnitude of the breakup cross section. These unrealistic values are due to a lack of absorption at small $b$s within the model. {As explained in Sec.~\ref{Sec2E},  the real parts of the potentials contribute to the imaginary parts of the phases~\eqref{eq14}--\eqref{eq15}. These contributions    decrease the absorption and lead to unrealistic values of the cross section.}

{To suppress these unphysical contributions of the real part of the potentials, we treat separately the real and  imaginary  parts of the nuclear potential.}
As shown in \Eq{eq16}, the S-DEA is applied to the Coulomb potential and the real part of the nuclear potentials. The imaginary part of the nuclear interaction is treated with the usual eikonal model. The corresponding distributions are plotted in  dash-dotted-dotted  black lines. For both  collisions with   carbon and lead targets, the S-DEA is accurate over the whole considered  energy range. Note that, the distribution of the  Coulomb-dominated breakup  still exhibits an unphysical increase at $E=12$~MeV, {caused by the imaginary part of the} Coulomb first-order  approximation~\eqref{eq14}. However, at these energies, the breakup cross section is negligible. There is no such increase in the nuclear-dominated collision since the Sommerfeld parameter $\eta$ and thus the Coulomb first-order-perturbation  approximation are smaller.

\begin{figure*}
	\center
	{\includegraphics[width=0.45\linewidth]{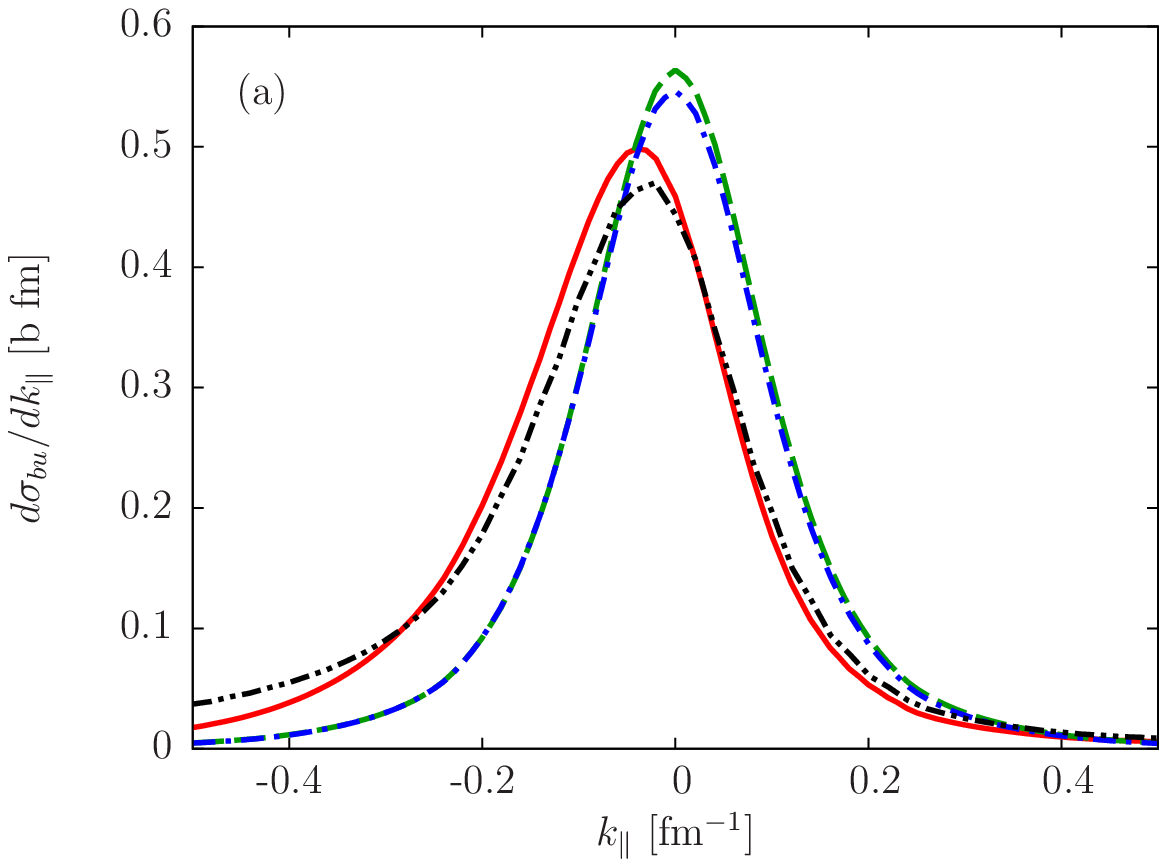}}
	\hspace{0.5cm}
		{\includegraphics[width=0.45\linewidth]{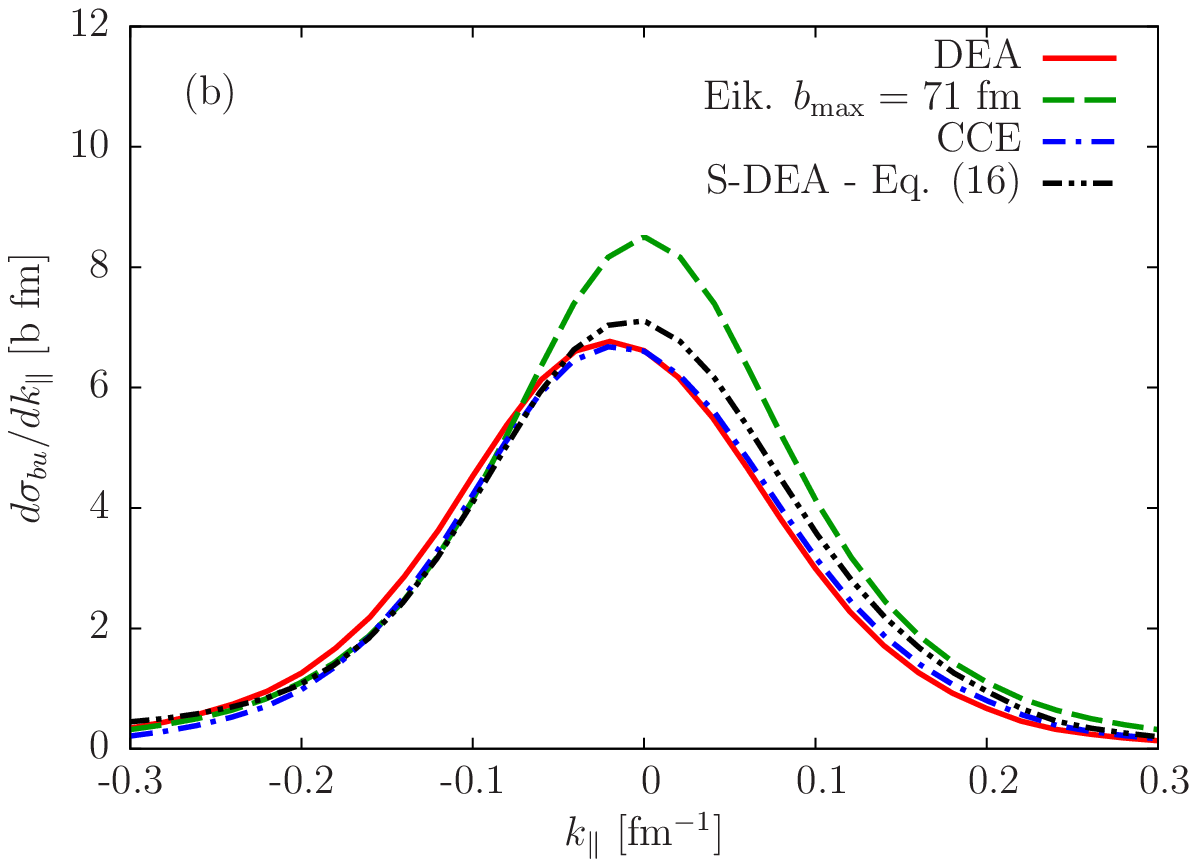}}	
	\caption{Diffractive breakup distribution of $^{11}\rm Be$ with (a)  $^{12}\rm C$ at $67A$~MeV  and (b) $^{208}\rm Pb$ at 69$A$~MeV as a function of $^{10}\rm Be$-$n$  parallel momentum. } \label{FigSbudkp}
\end{figure*}

We now consider in Fig.~\ref{FigSbudkp}(a) the breakup distribution as a function of the relative $^{10}\rm Be$-$n$ parallel momentum {[see Eq.~\eqref{eq18} below] for} the breakup of $^{11}\rm Be$ on $^{12}\rm C$ at $67A$~MeV\footnote{{We have noted that the magnitudes of the  parallel-momentum distributions in Figs. 4, 5, 8 and 9 of  Ref.~\cite{CBS08} are underestimated by a factor 2. }}. In this case, the eikonal model does not display a divergence since the nuclear interaction dominates. However, it overestimates the magnitude of the cross section and does not reproduce the asymmetry of the distribution. Indeed, the eikonal distribution is perfectly symmetric, due to the additional symmetry of the projectile across the  plane defined by ${\ve{s}}$ (see Sec.~\ref{Sec2C}).  The CCE cross section lies close to eikonal results, because its Coulomb correction~\eqref{eq8} is not significant for such nuclear-dominated reaction. On the contrary, the S-DEA leads to a more accurate distribution, enhancing both shape and magnitude. This result confirms that the asymmetry of the distribution is due to dynamical effects and shows that a first-order simplification of the DEA already improves significantly the distributions. Further analyses have shown that the $d5/2$ resonance does not impact the accuracy of the S-DEA. Similarly to the energy distribution, the S-DEA underestimates the DEA magnitude.

 The parallel-momentum distributions obtained for the collision of $^{11}\rm Be$ on $^{208}\rm Pb$  at 69$A$~MeV are plotted in Fig.~\ref{FigSbudkp}(b). As for the energy distributions, the eikonal model fails to describe the parallel-momentum distribution: the  magnitude and shape of the distribution are different from the DEA predictions.  On the contrary, the CCE lies  close to the DEA results and is  precise for this Coulomb-dominated breakup. The S-DEA  slightly overestimates the peak of the distribution and is too symmetrical compared to the DEA. Since this collision is dominated by the Coulomb interaction, this difference is due to the treatment of electric transitions within the CCE and S-DEA.
 
To understand the origin of the asymmetry, we consider the expression of the parallel-momentum observable, which reads
 		\beq
 		\lefteqn{\frac{d\sigma_{bu}}{dk_\parallel}=\frac{8\pi}{2j_0+1}
 			\sum_{m_0}\int_0^\infty b db {\int_{|k_\parallel|}^{k_{\rm max}}} \frac{dk}{k}
 			\sum_{\nu m}}\nonumber\\
 		&&\times \left|\sum_{lj}(lI m-\nu \nu|jm)
 		Y_l^{m-\nu}(\theta_k,0)S_{kljm}^{(m_0)}(b)\right|^2,
 		\eeqn{eq18}
 		where $\theta_k=\arccos (k_\parallel/k)$ is the colatitude of the
 		$c$-$n$ relative wave vector $\ve{k}$ after breakup, $l$ is the orbital angular momentum of the $c$-$n$ system, $j$ is the total angular momentum, resulting from
 		the composition of $l$ and the spin of the neutron $I$, $m$ is its projection and $\nu$ is the projection of the spin. The subscript $0$ corresponds to the quantum numbers of the ground state $\phi_0$. The partial breakup amplitude $S_{kljm}^{(m_0)}$ is  the matrix element of the operator~\eqref{eq8} [resp.~\eqref{eq16}] for the CCE (resp. for the S-DEA) between the initial state $\phi_0$ and the final breakup state (see Eq. (25) of Ref.~\cite{CBS08}). 
{The integration over $k$ is limited to $k_{\rm max}=0.7$~fm$^{-1}$ for the $\rm Pb$ target and $k_{\rm max}=1.4$~fm$^{-1}$ for the $\rm C$ target, which correspond respectively to $E_{\rm max}=11.26$~MeV and $E_{\rm max}=45$~MeV. These values are large enough to reach convergence.}

{This observable coherently sums the  partial breakup amplitudes and is therefore sensitive to their interferences.
With the property of the spherical harmonics
	\begin{equation}
	Y_{l}^{m-\nu}(\theta_k,0)=(-1)^{l+m-\nu} 	Y_{l}^{m-\nu}(\pi-\theta_k,0),
	\end{equation}
each partial-wave contribution to the parallel-momentum distribution differs by a phase $(-1)^l$ for two opposite $k_\parallel$ values, making opposite interferences between even $l$ and odd $l$ partial waves.}
In both models, the asymmetry of the distribution is caused by  the terms asymmetric in $z$,  contained in   the breakup amplitudes with odd $l+m-\nu$ as the ground state is in the $s$ wave.

	\begin{figure*}
		\center
		{\includegraphics[width=0.45\linewidth]{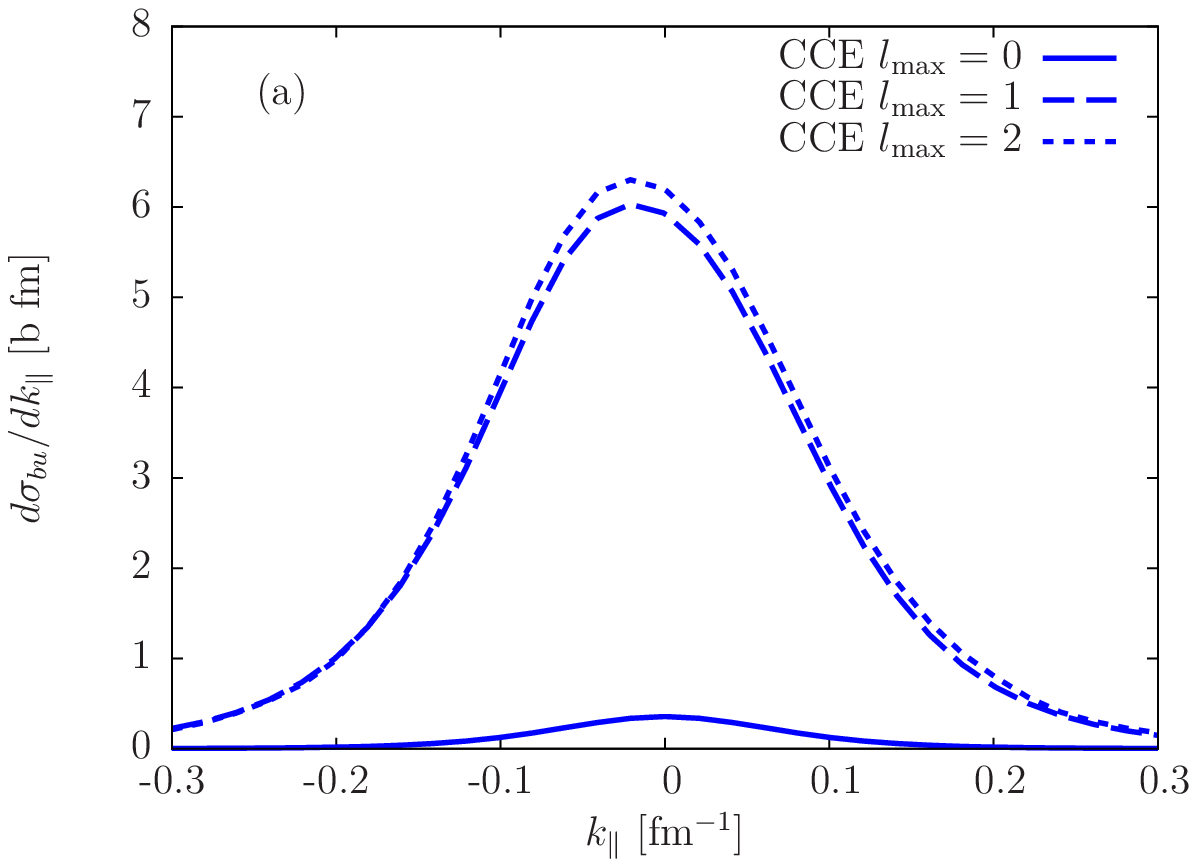}}
		\hspace{0.5cm}
		{\includegraphics[width=0.45\linewidth]{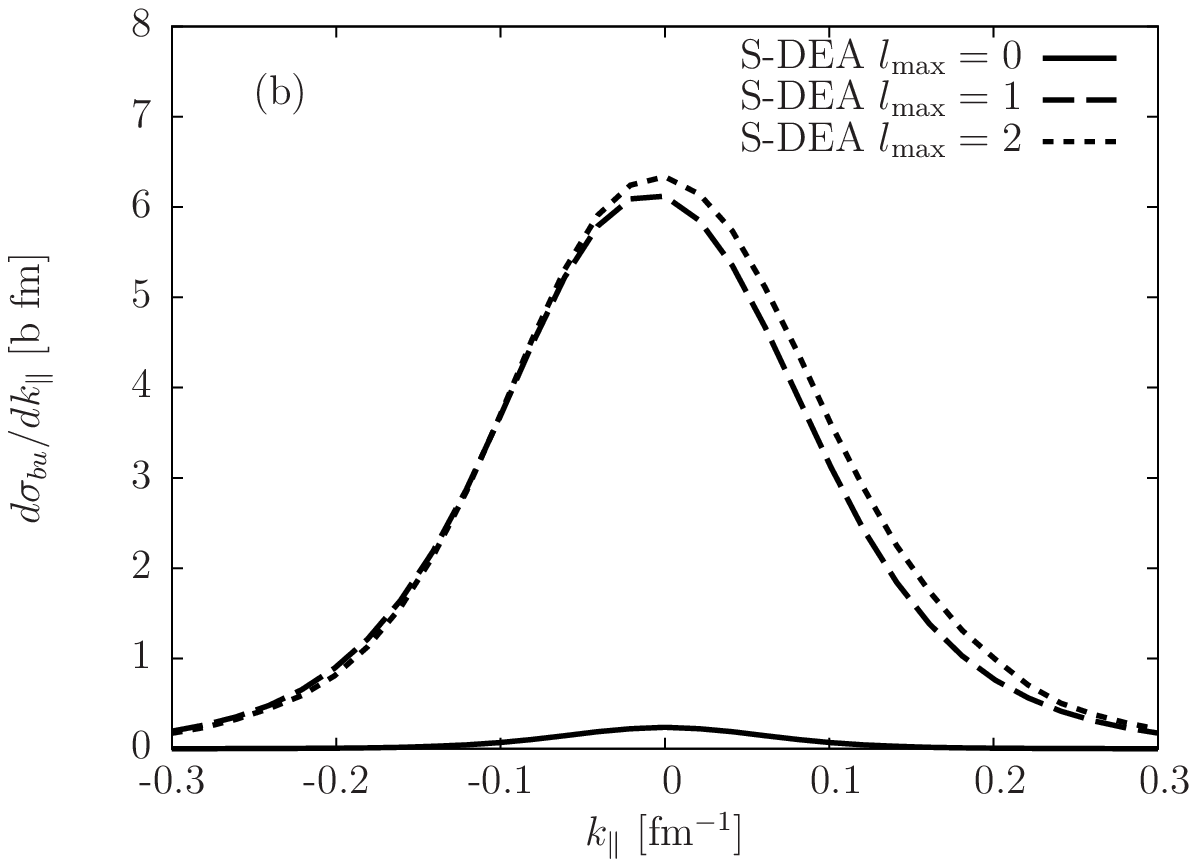}}	
		\caption{Coulomb contribution to  breakup  of $^{11}\rm Be$ with  $^{208}\rm Pb$ at 69$A$~MeV obtained with the (a) CCE and (b) S-DEA as a function of $^{10}\rm Be$-$n$  parallel momentum. A cutoff at $b_{\rm min}=10$~fm is used to simulate the nuclear absorption.} \label{FigSbudkpla}
	\end{figure*}
{For a pure Coulomb breakup}, only the imaginary part of $\chi^C_{\rm FO}$ [\Eq{eq14}] is responsible for the asymmetries of the CCE and S-DEA distributions.
In the CCE, this asymmetric term contributes {at only $l=1$ and not to all $l$ as in the S-DEA}, where $\chi^C_{\rm FO}$ appears in the exponential~\eqref{eq16}. To understand how this affects the shape of the parallel-momentum distributions, we plot in Fig.~\ref{FigSbudkpla} its Coulomb contribution for the breakup of $^{11}\rm Be$ with $^{208}\rm Pb$, obtained with (a) the CCE and (b) the S-DEA.  We use a cutoff at $b_{\rm min}=10$~fm from which we compute the cross sections to simulate the nuclear absorption. As expected, the {$l=0$ term} (solid lines) leads to  symmetric distributions for both the CCE and S-DEA. For both models, the asymmetry {first} arises from the interferences of the {$l=0$ and $l=1$} terms (dashed lines). Moreover, the magnitude of the distributions increases drastically when we add the {$l=1$} term, because this  reaction is dominated by E1 transitions.
	 
The main difference between the two models is the influence of the {$l=2$} term: in the CCE, it barely influences the shape of the distribution while in the S-DEA, it makes it more symmetric. {In the CCE, the {$l=2$} term depends on $\chi^C$ [\Eq{eq8}], which is symmetric. Since the collision is dominated by E1 transitions, it does not impact much the distribution.} In the S-DEA, the {$l=2$} term depends on $\chi_{{\rm FO}}^C$, and thus also contains an asymmetric {component. However, it contributes in an opposite way to the {$l=1$} term, i.e., more to positive $k_\parallel$ values. It} tends to diminish the asymmetry of the distribution. Therefore, each  {{partial wave $l$} influences the asymmetry in the S-DEA and the total effect reduces} the asymmetry due to the E1 transition.

{The asymmetry of the breakup distribution of $^{11}\rm Be$ with $^{12}\rm C$ in Fig.~\ref{FigSbudkp} obtained at the S-DEA can  also be analysed. In this case, the nuclear interaction dominates and  therefore the asymmetry is mainly influenced by the imaginary part of  $\chi^N_{\rm S-DEA}$~\eqref{eq17}. Contrary to the Coulomb case, this phase has two  imaginary exponentials, one depending on $-z$, and another one on $+z$. Therefore, ${\rm Im}\,\chi^N_{\rm S-DEA}$  contains two   asymmetric terms which have an opposite sign and their sum  depends on  $z$,  $b_{cT}$ and $b_{nT}$.  This complex dependence implies that  the partial waves with odd and even $l$ do not necessarily have opposite contributions to the asymmetry of the parallel-momentum distribution.  In this case, we have observed that in a pure nuclear calculation, both odd and even partial waves $l$ contribute to the asymmetry, i.e, push the distribution to negative $k_\parallel$.   Contrary to the Coulomb-dominated reaction, the asymmetry of the total distribution (in Fig.~\ref{FigSbudkpla}) is overestimated. This  analysis suggests that the S-DEA improves the descriptions of electric and nuclear transitions but still misses part of the dynamics of the reaction.}

In conclusion, the S-DEA with a separate treatment of the real and imaginary parts of the nuclear potential leads to  accurate energy distributions for both nuclear- and Coulomb-dominated breakups. We have also shown that the S-DEA improves the  parallel-momentum distribution for both reactions compared to the usual eikonal model. Moreover, the S-DEA reproduces better the asymmetry of the distribution of nuclear-dominated breakup compared to the CCE. It misses however part of the asymmetry of the Coulomb-dominated distribution, due to destructive interferences between different partial waves, decreasing the {asymmetry} of the distribution.

\section{Conclusions}\label{Conclusions}
To infer reliable information from breakup observables, one needs an accurate reaction model coupled with a realistic description of the collision. The eikonal model  has the advantage of being  cheap from a computational viewpoint, while keeping the quantal description of the collision. Unfortunately,  it does not treat well the Coulomb interaction, causing a divergence within the breakup matrix element.

We develop in this work a simplification to the DEA, which has a similar numerical cost as the usual eikonal approximation. This model  is based on an approximation of the solutions of the DEA wave functions. We make a second approximation by replacing the internal Hamiltonian by their eigenvalues within the breakup matrix element. This leads to an eikonal-like model, where the eikonal phases are replaced by the first-order-perturbation approximations. Compared to the  eikonal model, this model does not require any cutoff and {its wave functions tend to the first-order-perturbation theory}. These approximations are applied to only the real part of the potentials, leaving the imaginary part to the usual eikonal model. This avoids a lack of absorption and gives precise energy distributions.

The S-DEA also improves  the breakup distribution as a function of the  $^{10}\rm Be$-$n$ parallel momentum. Compared to the usual eikonal model, it lies closer to the DEA results.  For nuclear-dominated breakup, the S-DEA improves both the shape and the magnitude of the distribution, confirming that the asymmetry is due to dynamical effects during the collision. It is therefore an excellent model to describe both energy distribution and parallel-distribution observables of breakup with light targets. The S-DEA does however not reproduce the asymmetry of the parallel-momentum distribution of {this Coulomb-dominated reaction. The  contribution of higher multipoles {of the wave function} weakens the interferences caused by the dominant E1 transition.}

Since the S-DEA  elegantly solves  the Coulomb divergence within the eikonal model and significantly enhances  its accuracy while keeping a small numerical cost, it would be interesting to study its extension  to three-body projectiles, such as two-neutron halo nuclei. As there are still some differences with the DEA, higher-order approximations of the DEA solutions could be studied. A first step could be to improve the crude approximation of the exponential operators of \Eq{eq11} within the breakup matrix elements. This could be done by expanding the wave function onto a basis of eigenstates and pseudostates for the $c$-$n$ continuum of the internal Hamiltonian (as in CDCC~\cite{Kam86,YOMM12}). The  matrix elements could then be obtained  through a diagonalization method~\cite{AW60,AW75,BH73} and an interpolation of the energies.

\begin{acknowledgements}
	We thank P. Capel for interesting discussions on this topic.	C.~Hebborn acknowledges the support of the Fund for Research Training in Industry and Agriculture (FRIA), Belgium.  This project has received funding from the European Union’s Horizon 2020 research	and innovation program under grant agreement 	No 654002. This work was supported by the Fonds de la Recherche Scientifique - FNRS under Grant Number 4.45.10.08.
\end{acknowledgements}

\bibliographystyle{apsrev}
\bibliography{HBC_ACE}
\end{document}